\documentclass[aps,prd,twocolumn,superscriptaddress]{revtex4}
\usepackage{epsfig}
\usepackage{amsmath}
\usepackage{graphicx}
\usepackage[mathcal]{eucal}

\usepackage{graphicx}
\newcommand{\ben}{\begin{equation}}
\newcommand{\een}{\end{equation}}
\newcommand{\bea}{\begin{eqnarray}}
\newcommand{\eea}{\end{eqnarray}}

\newcommand{\eqn}[1]{Eq.~(\ref{#1})}

\newcommand{\sm}{{SM}}

\newcommand{\dred}{{DRED}}
\newcommand{\rg}{{RG}}

\def\GeV{\hbox{GeV}}

\def\nn{\nonumber\\}
\def\DRED{\ifmmode{{\rm DRED}} \else{{DRED}} \fi}
\def\DREDD{\ifmmode{{\rm DRED}'} \else{${\rm DRED}'$} \fi}
\def\NSVZ{\ifmmode{{\rm NSVZ}} \else{{NSVZ}} \fi}

\def\pa{\partial}

\begin{document}

\title{The quadratic divergence in the Higgs mass revisited}

\author{D.~R. Timothy Jones}\email{drtj@liverpool.ac.uk}
\affiliation{Department of Mathematical Sciences, University of
Liverpool, Liverpool L69 3BX, UK}

%\date{Sept 8 2013}

%\preprint{\vbox{\hbox{LTH 984}}}

\begin{abstract}
Old and new calculations of the Higgs mass quadratic divergence are compared. 
\end{abstract}

%\begin{flushright}
%\vspace{-36pt} 
%%
%{\hfill LTH 984}
%\end{flushright}

\maketitle

%%%%%%%%%%%%%%%%%%%%%%%%%%%%%%%%%%%%%%%%%%%%%
\section{Introduction}

The standard model (\sm) Higgs-like particle of mass $125\GeV$ recently discovered at
the LHC~\cite{:2012si,Chatrchyan:2012tx} has 
resulted in a revival of interest in the old Veltman 
observation~\cite{Veltman:1980mj}
that it is possible to arrange cancellation of the quadratic divergence 
in the Higgs mass by imposing a certain relation upon the coupling 
constants of the theory~\cite{Hamada:2012bp}-\cite{Bian:2013xra}. 
In the notation of 
\cite{Hamada:2012bp}, 
the quadratic divergence at one loop is 
proportional to 
\ben\label{eq:quad1}
Q_1 = \lambda + \frac{1}{8}{g'}^2 + \frac{3}{8}g^2 -y_t^2. 
\een
Veltman, in fact, expressed the relation in terms of the particle masses, thus:
\ben
Q'_1= 2Q_1 v^2 = m_H^2 + 2m_W^2 +m_Z^2 -4m_t^2, 
\label{eq:masses}
\een
whereas \cite{Alsarhi:1991ji} opted to use mass ratios.
In the notation of \cite{Alsarhi:1991ji}:
\ben
\Delta_1 = Q'_1/m_W^2 = H + 3 + \tan^2\theta_W - 4T
\een 
where $m_W^2 = \frac{1}{4}g^2v^2$, $m_Z^2 = \frac{1}{4}(g^2 + {g'}^2)v^2, 
m_H^2 = 2\lambda v^2, 
m_t^2 = \frac{1}{2}y_t^2 v^2, H = m_H^2/m_W^2, T = m_t^2/m_W^2$, and $v$ is the Higgs vev. 
Veltman, I believe, thought of the relation as existing for the {\it
physical\/} masses  of the particles, and in his original paper opted to
perform the calculation in the  broken phase of the theory  (although
the symmetric phase calculation is much simpler).
Requiring $Q'_1=0$ predicts $m_H \approx 315\GeV$, clearly at odds with
the recent observations. 

Now if it really was in terms of physical couplings, \eqn{eq:masses}\ 
would be renormalisation group invariant. 
However \eqn{eq:quad1}, expressed as it is in terms of renormalised
couplings, is clearly renormalisation scale dependent,
and recent interest in it has centered on the effect of running
$Q_1$ up to higher energies and perhaps matching it on to an
underlying supersymmetric theory at some scale~\cite{Masina:2013wja},
\cite{Casas:2004gh}.  The
observation~\cite{Casas:2004gh},\cite{Hamada:2012bp} that $Q_1$
changes sign at some high scale (the value of which scale being quite
sensitive to the precise value of the top mass)  has led to the
remarkable suggestion~\cite{Jegerlehner:2013cta} that this sign change
is actually the trigger for electroweak symmetry breaking.
 
In fact the issue of the scale dependence of $Q_1$ was considered
in general theories and in the particular case of the SM many years
ago~\cite{Alsarhi:1991ji}-\cite{Einhorn:1992um}. This work included the
observation that in a Yukawa-scalar non-gauge theory, there exists a
intriguing relationship between the scale dependence of $Q_1$ and
the leading quadratic divergence at the two loop level. In fact,
requiring $Q_1$ to be both zero and scale independent to leading
order in the $\beta$-functions leads to precisely the same condition as
requiring the 2-loop leading quadratic divergence to vanish! 

In \cite{Jack:1989tv}, the {\it leading\/} quadratic  divergence
at $L$ loops was defined in the context of regularisation by dimensional
reduction  (\dred)~\cite{Siegel:1979wq}, \cite{Capper:1979ns} as the
residue of the pole at $d=4-2/L$ in the IR-regulated 2-particle
amplitude. This definition corresponds, in fact, to associating the {\it
leading\/} quadratic divergence at two loops with the (IR regulated)
integral
\ben
I_2 = \int \frac {d^d k \, d^d q}{k^2 q^2 (k+q)^2}
\een
which is precisely what is done in \cite{Hamada:2012bp}.
At two loops one also encounters
 \ben
I_1 = \int \frac {d^d k \, d^d q}{(k^2)^2 q^2 }
\een
which has a pole at $d=2$ and is cancelled by the one-loop counter-term 
insertion contribution.  

In \cite{Hamada:2012bp}, a calculation of the two loop quadratic
divergence in the Higgs mass is presented, and the coefficient of $I_2$
is found to be  proportional to $Q_2$ where 
\bea 
Q_2 &=& -(9y_t^4+y_t^2(-\frac{7}{12}g'^2 +\frac{9}{4}g^2 -16g_3^2)+\frac{77}{16}g'^4\nn
&+& \frac{243}{16}g^4 +\lambda (-18y_t^2+3g'^2+9g^2)-10\lambda^2).
\eea 

It appears the authors were unaware of the previous calculation of  the
same quantity\footnote{and, indeed, this quantity in a general
renormalisable  theory\cite{Jack:1989tv}.} (using \dred) of
\cite{Alsarhi:1991ji}, where the result found was proportional to
$\Delta_2$, given by 
\bea\label{eq:delta2}
\Delta_2 &=& 
\frac{9}{2}H^2 +27HT-54T^2-9H(3+\tan^2\theta_W)\nn
&-&T(27-7\tan^2\theta_W-s)\nn
&+&\frac{189}{2}+45\tan^2\theta_W + \frac{261}{2}\tan^4\theta_W,
\eea
where $s =192g_3^2/g^2$.

Reducing $\Delta_2$ to the same notation as $Q_2$ we obtain
 \bea
m_W^4 \Delta_2 &=& -\frac{3}{2}(9y_t^4
+y_t^2(-\frac{7}{12}g'^2 +\frac{9}{4}g^2-16g_3^2)\nn 
&-&\frac{87}{16}g'^4 -\frac{63}{16}g^4 - \frac{15}{8}g^2 g'^2\nn
&+&\lambda (-18y_t^2+3g'^2+9g^2)-12\lambda^2),
\label{eq:delta2sm}
\eea
and we see that most terms agree. (The overall factor is not significant; in 
\cite{Alsarhi:1991ji}-\cite{AlSarhi:1990np} 
we were concerned with seeking theories 
{\it without\/} quadratic divergences). However the 
$\lambda^2, g^4, g'^4$ and $g^2 g'^2$ terms do not agree; in both magnitude 
and sign in the case of the 
$ g^4, g'^4$ terms. The disagreement was noted in \cite{Jegerlehner:2013cta}, the author 
of which opted to believe the result of \cite{Hamada:2012bp}.

Note that the result of~\cite{Hamada:2012bp} has no $g^2 g'^2$ term. 
On this particular point we can easily see, I believe, that \cite{Hamada:2012bp} is 
incorrect as follows. 

The calculations of \cite{Hamada:2012bp} were done in the Landau gauge, 
in which gauge, as they remark, it is easy to see that graphs of the 
general form of Fig.~1 do not contribute. 
\begin{figure}[ht]
%\FIGURE{
\begin{center}
\includegraphics[scale=0.4]{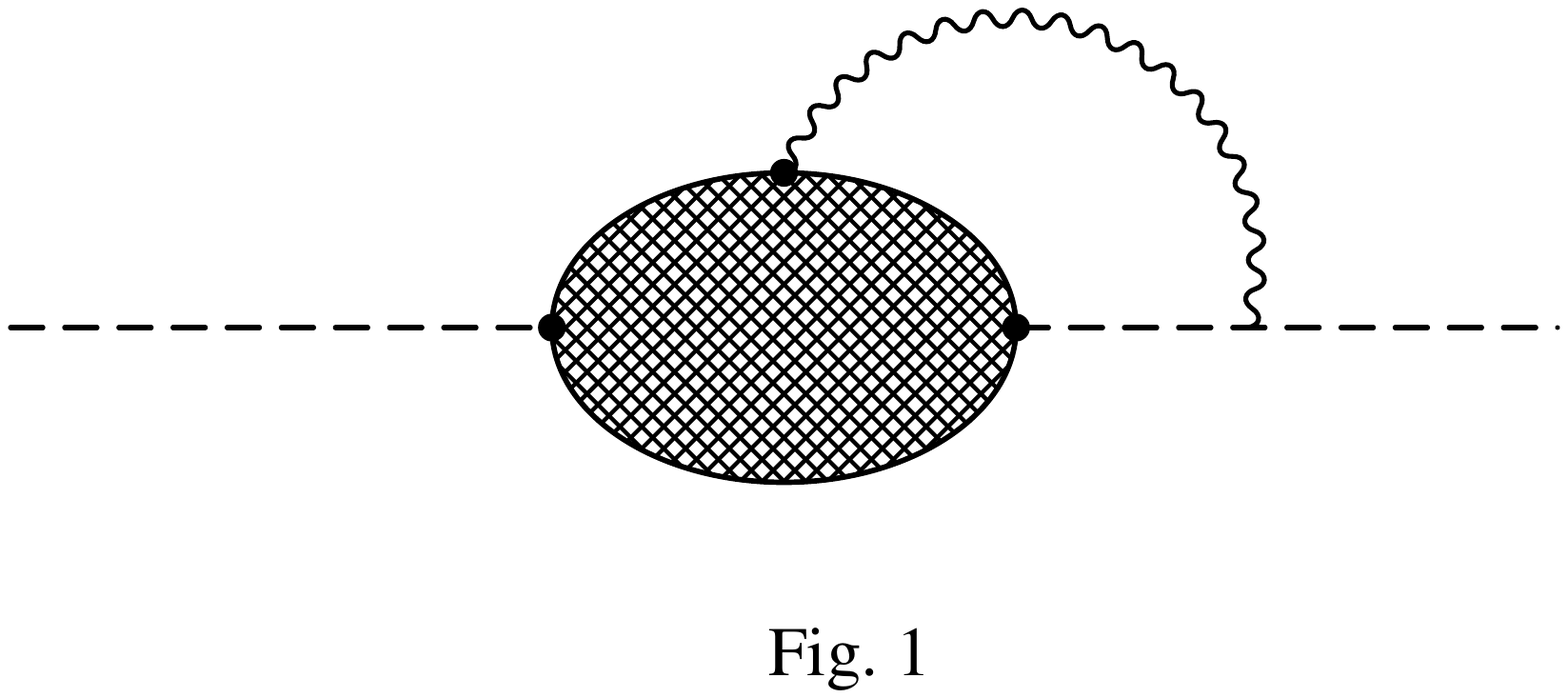}
\caption{A class of graphs free of quadratic divergences
\label{zerographs}
}
\end{center}
%}
\end{figure}
In the Landau gauge there is, however, one graph that {\it does\/} 
give rise to a $g^2 g'^2$ term,
shown in Fig.~2. 
\begin{figure}[ht]
%\FIGURE{
\begin{center}
\includegraphics[scale=0.4]{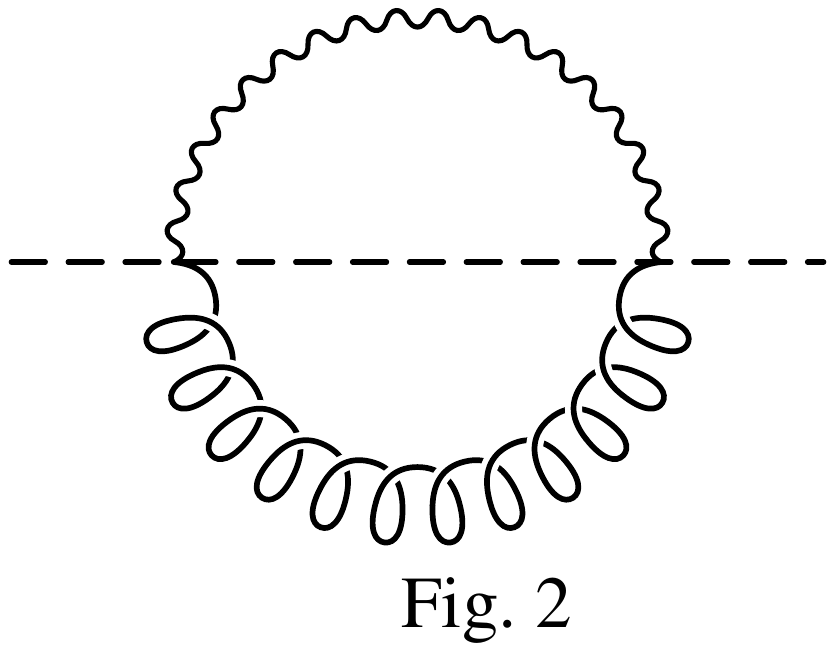}
\caption{Non-zero contribution proportional to $g^2g'^2$.
\label{nonzerograph} 
}
\end{center}
%}
\end{figure}

I have calculated the graph shown in Fig.~2 in the Landau gauge, and obtained a result
in agreement with \eqn{eq:delta2}. It seems to me likely that the
authors of \cite{Hamada:2012bp} have inadvertently omitted this
graph. 

With regard to the remaining discrepancies, the difference in the 
$\lambda^2$ terms presumably results from an error by one group or the
other. For the $g^4, g'^4$ terms, two issues arise. The first is gauge
invariance; I am not aware of a proof that the whole result is gauge
invariant, but I believe it is. The fact that I have obtained the same
result for the  $g^2g'^2$ term using the Landau gauge as that of
\cite{Alsarhi:1991ji} (where the  calculations were performed in a
background Feynman gauge,  using configuration space methods) is some
evidence for this.  The second issue arises from the the fact that using
\dred, the $\epsilon$-scalars  peculiar to that scheme {\it themselves\/} develop a
one loop self-energy quadratic divergence. As described in
\cite{Alsarhi:1991ji}, this leads to a breakdown in the
relationship between the leading two loop divergence $Q_2$ and the
quantity 
\ben A_{11} = \beta_{\lambda_i}^{(1)}.\frac{\pa}{\pa\lambda_i}
Q_1  - Q_1.\frac{\pa}{\pa\lambda_i}\beta_{\lambda_i}^{(1)} 
\een 
that, as mentioned above, had been observed in non-gauge theories. 
It would thus have been very interesting had the result of 
\cite{Hamada:2012bp} for $Q_2$ agreed with $A_{11}$ but it does not.  
In any
event, I believe that using \dred\ and identifying the $d=3$ pole  is
equivalent to the  procedure of \cite{Hamada:2012bp}. 

My confidence in the result of~\cite{Alsarhi:1991ji} relies on the
general results Eqs.~(3.5),(3.8) given there and the \rg\ check on the
reduction to the SM case described in the appendix of that reference. In
this context, however, I should remark that there is a typo in Eq.~(4.3)
of the published version of that reference, which should read
\bea\label{eq:a11}
A_{11} &=& 
\frac{9}{2}H^2 +27HT-54T^2-9H(3+\tan^2\theta_W)\nn
&-&T(27-7\tan^2\theta_W-s)
+\frac{21}{2}+45\tan^2\theta_W\nn
&+& \frac{109}{2}\tan^4\theta_W.
\eea
Note that Eq.~(4.5) of~\cite{Alsarhi:1991ji}, which is obtained by substituting 
$\Delta_1 = 0$ from Eq.~(4.1) in Eq.~(4.3), is in fact correct. 
From \eqn{eq:a11} we obtain 
\bea
m_W^4 A_{11} &=& -\frac{3}{2}(9y_t^4
+y_t^2(-\frac{7}{12}g'^2 +\frac{9}{4}g^2-16g_3^2)\nn 
&-&\frac{109}{48}g'^4 -\frac{7}{16}g^4 - \frac{15}{8}g^2 g'^2\nn
&+&\lambda (-18y_t^2+3g'^2+9g^2)-12\lambda^2).
\label{eq:a11sm}
\eea
The difference between $A_{11}$ and $\Delta_2$ was, as we indicated 
above, associated by \cite{Alsarhi:1991ji} with the $\epsilon$-scalar
self-energy  component of the diagrams shown in Fig.~3 (in fact only 
Fig.~3b contributes). It is easy to check that the difference between
\eqn{eq:delta2sm}  and \eqn{eq:a11sm} above is consistent with
Eq.~(3.9)\ of \cite{Alsarhi:1991ji}.

\begin{figure*}[!t]
%\FIGURE{
\begin{center}
\includegraphics[scale=0.4]{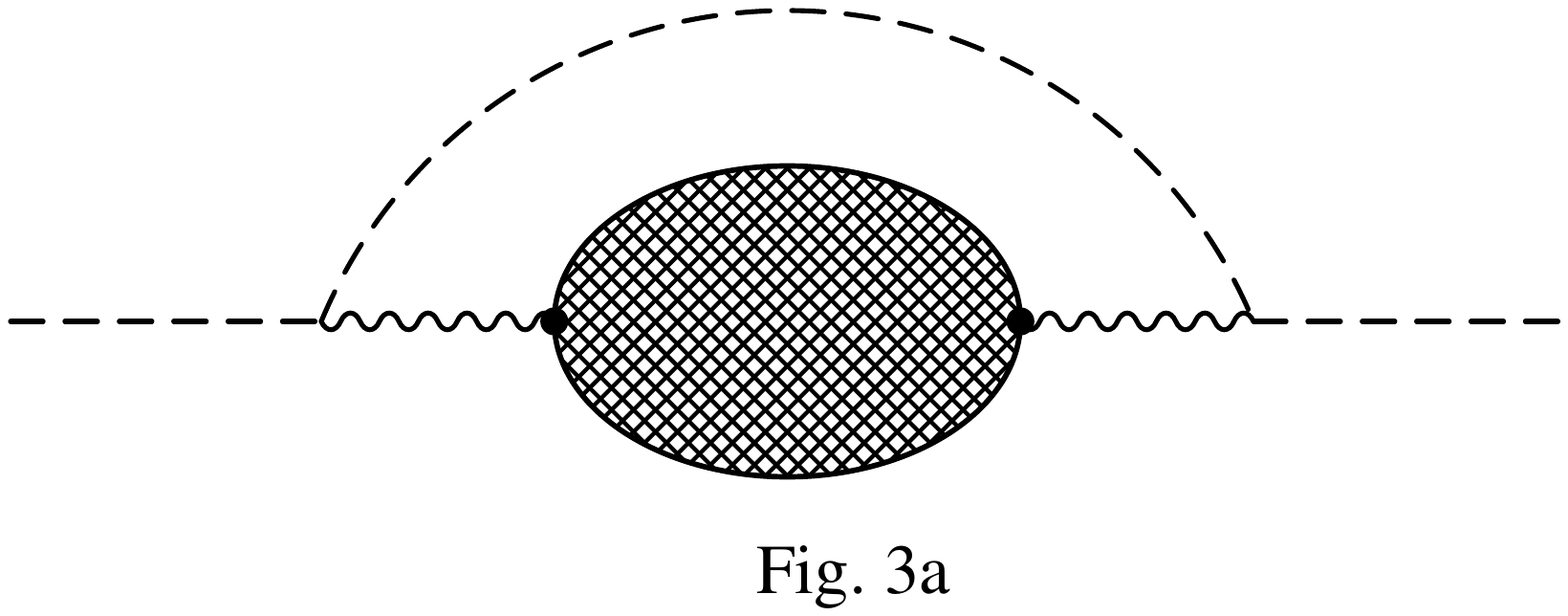}
\includegraphics[scale=0.4]{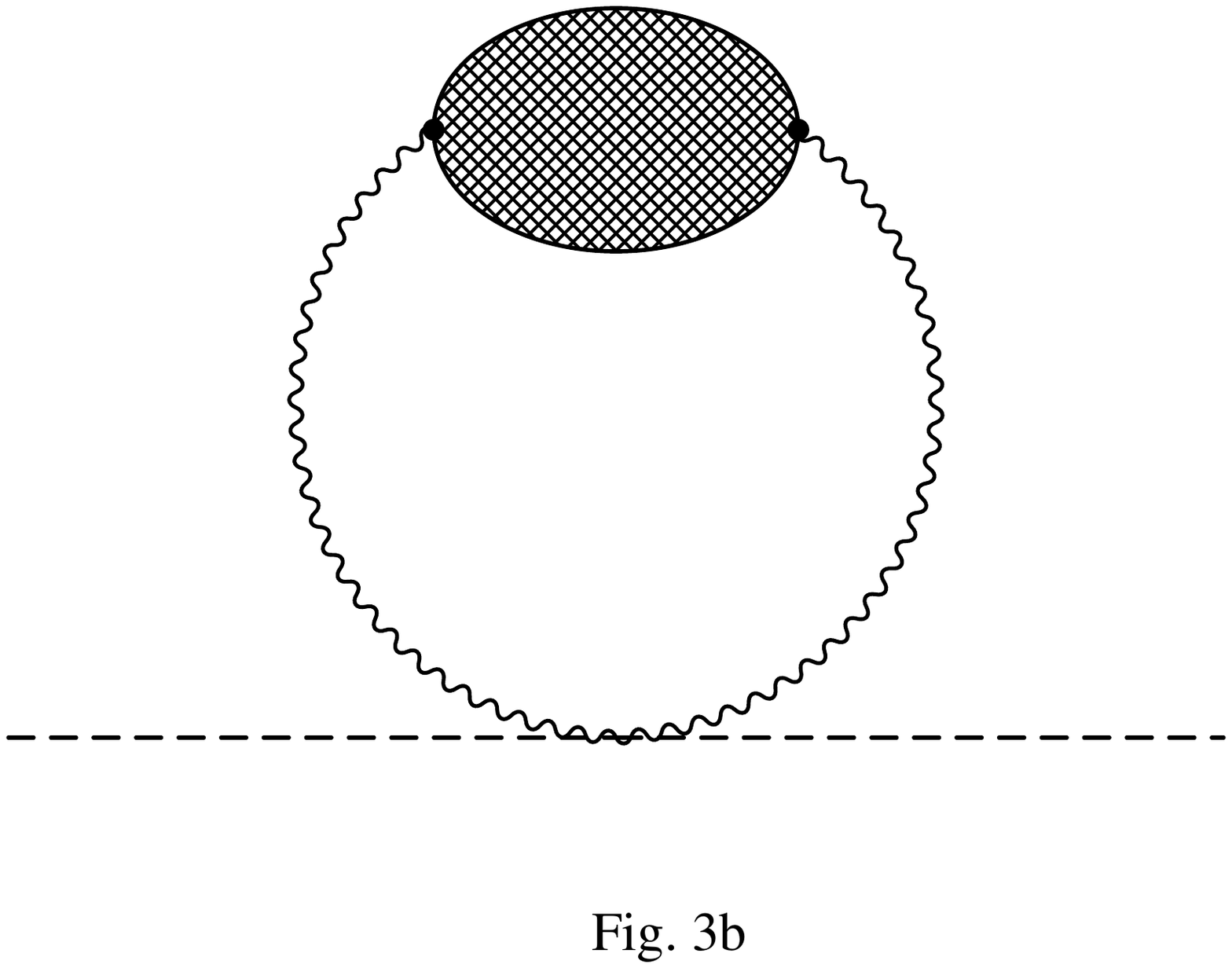}
\caption{A class of graphs with the $\epsilon$-scalar self energy
\label{epsilongraphs}
}
\end{center}
%}
\end{figure*}

With a {\it physical\/} cut-off for the quadratic divergence, it is
reasonable to argue~\cite{Hamada:2012bp} that, away from $Q_1 = 0$,
the effect of the two loop quadratic divergence  $Q_2$ is small 
compared to that of $Q_1$. Therefore the disagreements I have
indicated above will not have much impact on the thrust of the arguments
presented in  \cite{Hamada:2012bp}-\cite{Bian:2013xra}, although it may 
well change the scale at which the {\it total\/} quadratic divergence 
reaches zero by an appreciable amount. For possible future applications 
it is as well to clarify which of the two calculations discussed here is
correct.

\acknowledgments

This research was supported in part by the Science and Technology
Research Council [grant number ST/J000493/1]. 
I thank Marty Einhorn for drawing my attention to \cite{Hamada:2012bp}, 
and Ian Jack for reintroducing me to FeynDiagram.

\end{document}